\input epsf.tex    

\documentstyle[11pt]{article}

\newcommand{\bmat}{\left(\begin{array}}
\newcommand{\emat}{\end{array}\right)}

\def\yzero{\smash{\hbox{$y\kern-4pt\raise1pt\hbox{${}^\circ$}$}}}

\def\beq{\begin{equation}}
\def\eeq{\end{equation}}
\def\beqa{\begin{eqnarray}}
\def\eeqa{\end{eqnarray}}

\def\-{\hphantom{-}}
\def\ov{\overline}
\def\s2{\frac{1}{2}}

\def\beq{\begin{equation}}
\def\eeq{\end{equation}}
\def\beqa{\begin{eqnarray}}
\def\eeqa{\end{eqnarray}}
\def\tr{{\rm tr \,}}
\def\Tr{{\rm Tr \,}}

\def\diag{{\rm diag \,}}
\def\IF{\relax{\rm I\kern-.18em F}}
\def\II{\relax{\rm I\kern-.18em I}}

\def\cp{{\cal P}}
\def\IC{\bf C}
\def\IZ{\bf Z}
\def\IR{\bf R}

\def\IP{\bf P}
\def\IT{\bf T}
\def\IM{\bf M}
\def\IX{\bf X}
\def\z2z2{$\IC^3/(\IZ_2\times\IZ_2)$}

\def\id{{\bf 1}}

\def\Dsl{\,\raise.15ex\hbox{/}\mkern-13.5mu D} 

 \def\cp#1{\relax\ifmmode {\IP\kern-2pt{}_{#1}}\else $\IP\kern-2pt{}_{#1}$\=fi}
\newcommand{\drawsquare}[2]{\hbox{%
\rule{#2pt}{#1pt}\hskip-#2pt
\rule{#1pt}{#2pt}\hskip-#1pt
\rule[#1pt]{#1pt}{#2pt}}\rule[#1pt]{#2pt}{#2pt}\hskip-#2pt
\rule{#2pt}{#1pt}}

\newcommand{\fund}{\raisebox{-.5pt}{\drawsquare{6.5}{0.4}}}
\newcommand{\Ysymm}{\raisebox{-.5pt}{\drawsquare{6.5}{0.4}}\hskip-0.4pt%
        \raisebox{-.5pt}{\drawsquare{6.5}{0.4}}}
\newcommand{\Yasymm}{\raisebox{-3.5pt}{\drawsquare{6.5}{0.4}}\hskip-6.9pt%
        \raisebox{3pt}{\drawsquare{6.5}{0.4}}}
\newcommand{\antifund}{\overline{\fund}}

\begin{document}

\makeatletter \@addtoreset{equation}{section} \makeatother
\renewcommand{\theequation}{\thesection.\arabic{equation}}

\pagestyle{empty}
\vspace*{.5in}
\rightline{FTUAM-03/24, IFT-UAM/CSIC-03-45}
\rightline{\tt hep-th/0311250}
\vspace{1.5cm}
 
\begin{center}
\LARGE{\bf Chiral 4d String Vacua with D-branes and Moduli Stabilization 
\normalsize{\footnote{Contribution to the Proceedings of the Tenth Marcel 
Grossmann Meeting on General Relativity, Rio de Janeiro, July 20-26, 
2003}} \\[10mm]}

\medskip

\large{Juan F. G. Cascales$^1$, Angel M. Uranga$^2$} \\
{\normalsize {\em $^1$ Departamento de F\'{\i}sica Te\'orica C-XI \\
and Instituto de F\'{\i}sica Te\'orica, C-XVI \\
Universidad Aut\'onoma de Madrid \\
Cantoblanco, 28049 Madrid, Spain \\
{\tt juan.garcia@uam.es}\\
$^2$ IMAFF and \\
 Instituto de F\'{\i}sica Te\'orica, C-XVI \\
Universidad Aut\'onoma de Madrid \\
Cantoblanco, 28049 Madrid, Spain \\
{\tt angel.uranga@uam.es} \\[2mm]}}

\end{center}

\smallskip

\begin{center}
\begin{minipage}[h]{14.5cm}
{\small

We discuss type IIB orientifolds with D-branes, and NSNS and RR field 
strength fluxes, with D-brane sectors leading to open string spectra with 
non-abelian gauge symmetry and charged chiral fermions. The closed string 
field strengths generate a scalar potential stabilizing most moduli. Hence 
the models combine the advantages of leading to phenomenologically 
interesting (and even semirealistic) chiral open string spectra, and of 
stabilizing the dilaton and most geometric moduli. We describe the 
explicit construction of two classes of non-supersymmetric models on 
$\IT^6$ and orbifolds/orientifolds thereof, with chiral gauge sector 
arising from configurations of D3-branes at singularities, and from 
D9-branes with non-trivial world-volume magnetic fields. The latter 
examples yield the chiral spectrum of just the Standard Model.}
\end{minipage}
\end{center}

\newpage                                                        


\setcounter{page}{1} \pagestyle{plain}
\renewcommand{\thefootnote}{\arabic{footnote}}
\setcounter{footnote}{0}

\section{Introduction}

One of the most remarkable features of string theory is that, despite its
complexity, it admits vacua with low-enery physics surprisingly close to
the structure of observed particles and interactions. In particular there
exist by now several classes of constructions (e.g. heterotic 
compactifications, type II models with D-branes at singularities, 
intersecting D-branes, compactifications of Horava-Witten theory, etc) 
leading to four-dimensional gravitational and non-abelian gauge 
interactions, with charged chiral fermions. Within each class, particular 
explicit models with spectrum very close to that of the (Minimal 
Supersymmetric) Standard Model have been constructed. On the other hand, a 
generic feature of all these constructions, is the existence of a (very 
often large) number of moduli, which remain massless in the construction, 
unless some supersymmetry breaking mechanism is proposed. Even in the 
latter case, one often encounters runaway potentials for some moduli, with 
no minima at finite distance in moduli space. From this 
viewpoint these models are relatively far from describing physics similar 
to the observed world.

Recently, it has been shown that, in the setup of Calabi-Yau
compactification of type II string theory (or also M-theory), there exists a
natural mechanism which stabilizes most moduli of the compactification.
This is achieved by considering compactifications with non-trivial
field strength fluxes for closed string $p$-form fields. This proposal
has already been explored in different setups, leading to large classes
of models with very few unstabilized moduli. Hence this mechanism is one of
the most interesting recent insights to address the long-standing problems
of moduli in phenomenological string models. Unfortunately, compactifications 
with field strength fluxes have centered on simple models, which lead to 
uninteresting gauge sectors, from the phenomenological viewpoint. In
particular, the class of models studied are naturally non-chiral, since
the corresponding gauge sectors arise from too simple stacks of parallel
D-branes.

Our purpose in the present paper is to construct models which combine
the interesting features from the above two approaches. In fact, we 
construct string compactifications with interesting 4d chiral gauge 
sectors and flux stabilization of (most) moduli. The models are based on 
introducing NSNS and RR 3-form fluxes in compactifications of type IIB 
theory with D-branes. We discuss two classes of models, exploting 
different mechanisms to lead to chirality in the D-brane open string 
sector. In the first class, we construct models with D3-branes located at 
orbifold singularities, and chirality arises due to the orbifold 
projection. In the second class, we construct models with 
magnetised D9-branes, namely spacetime filling D9-branes with non-trivial 
gauge bundles on their world-volume, and chirality arises from a non-zero 
index of the Dirac operator in the Kaluza-Klein compactification of the 
charged 10d fermions (this mechanism is related, in the absence of 
fluxes, to the appearance of chirality at D-brane intersections via 
mirror symmetry). Although the complete models turn out to be 
non-supersymmetric, we discuss the stability properties of 
the resulting models.

For simplicity we center on constructions on $\IT^6$ and 
orbifolds/orientifolds thereof. The rules we describe are however quite 
general and we expect that the techniques and our new observations are 
useful in constructing other models, supersymmetric or not, in these and 
other geometries.

\smallskip

The paper is organized as follows. In Section \ref{fluxrev} we 
review the construction of toroidal orientifold compactifications with 
NSNS and RR 3-form fluxes. In section \ref{philosophy} we discuss
diverse classes of D-brane configurations leading to chiral gauge sectors, 
and issues arising in the possible introduction of 3-form fluxes for them.
In Section \ref{zthree} we describe the construction of models with 
D3-branes at singularities and NSNS and RR 3-form fluxes, and provide an 
explicit example with an open string spectrum yielding a 3-family $SU(5)$ 
GUT. In Section \ref{magnetised}, we describe the construction of models 
with D-branes with world-volume magnetic fields. We construct explicit 
models with 3-form fluxes and configurations of magnetised D-branes, with 
an open string spectrum yielding just the Standard Model, with three 
fermions families.

These constructions provide the first examples in a presumably rich and 
interesting class of realistic models. In Section \ref{final} we make our 
final comments.

This work elaborates over techniques and examples in \cite{cascur}, see 
also \cite{blt} for models closely related to those of section 
\ref{magnetised}.
 
\section{Review of fluxes}
\label{fluxrev}

Compactifications of type II theories (or orientifolds thereof) with
NSNS and RR field strength fluxes have been considered, among others, in
\cite{fluxes,dgs,moreflux,gkp,fp,kst,tt,ferrara,kors}. In this section we 
review properties of type IIB compactifications with 3-form fluxes.

\subsection{Consistency conditions and moduli stabilization}

Type IIB compactification on a Calabi-Yau threefold
\footnote{We should clarify that due to the flux backreaction, the metric 
is not the Ricci-flat metric in the Calabi-Yau, but rather conformal to it 
due to a non-trivial warp factor (sourced by the fluxes and objects in the 
background) \cite{dgs,gkp}.} $X_6$ with
non-trivial NSNS and RR 3-form field strength backgrounds $H_3$, $F_3$
have been extensively studied. In particular the analysis in 
\cite{dgs,gkp} provided, in a quite general setup, the consistency 
conditions such fluxes should satisfy. They must obey the Bianchi identities
\beqa
dF_3=0 \quad dH_3=0
\eeqa
and they should be properly quantized, namely for any 3-cycle
$\Sigma\subset {\bf X}_6$
\beqa
\frac{1}{(2\pi)^2\alpha'} \, \int_{\Sigma} F_3 \, \in {\IZ} \quad ; 
\quad
\frac{1}{(2\pi)^2\alpha'} \, \int_{\Sigma} H_3 \, \in {\IZ}
\eeqa
The fluxes hence define integer 3-cohomology classes in $H^3(X_6,\IZ)$.

A subtlety in flux quantization in toroidal orientifolds was noticed in 
\cite{fp,kst}. Namely, if flux integrals along some 3-cycle are integer 
but odd, consistency requires the corresponding 3-cycle to pass through an 
odd number of exotic O3-planes. For simplicity we restrict to the case 
where all flux integrals are even integers.

An important observation \cite{dgs,gkp} is that, to allow for non-trivial 
fluxes in compactifications to 4d Minkowski space, it is crucial to 
include orientifold 3-planes in the compactification, so we consider type 
IIB orientifolds with these objects. The simplest way to understand the 
need of these objects is to notice the type IIB supergravity 
Chern-Simons coupling
\beqa
\int_{M_4\times X_6}\, H_3 \wedge F_3 \wedge C_4
\label{cs}
\eeqa
where $C_4$ is the IIB self-dual 4-form gauge potential. This coupling 
implies that upon compactification the flux background contributes to a 
tadpole for $C_4$, with positive coefficient $N_{\rm flux}$ (in D3-brane 
charge units). Moreover, fluxes contribute positively to the energy of the 
configuration, due to the 2-form kinetic terms. The only way to cancel 
these tadpoles is to introduce O3-planes, objects with negative RR 
$C_4$-charge and negative tension, to cancel both the RR tadpole and also
to compensate the vacuum energy of the configuration. 

Having O3-planes in the configuration, it is natural to consider the 
possibility of adding $N_{Q_3}$ explicit D3-branes as well. The RR tadpole 
cancellation constraint hence reads
\beqa
N_{Q_3}\, + \, N_{\rm flux}\, + Q_{O3}=\, 0
\eeqa
We normalize charge such that a D3-brane in covering space has charge 
$+1$. With this convention an O3-plane has charge $-1/2$, and 
\beqa
\label{Nflux}
N_{\rm flux}\, = \, \frac{1}{(4\pi^2\alpha')^2}\, \int_{X_6}\, H_3\wedge 
F_3 \,=\, \frac{1}{(4\pi^2\alpha')^2} \, \frac{i}{2\phi_I}\,
\int_{X_6} \, G_3\wedge {\ov G}_3
\eeqa
where $\phi_I$ is the imaginary part of the IIB complex coupling $\phi
=a+i/g_s$, and
\beqa
G_3=F_3-\phi H_3
\label{gflux}
\eeqa
Finally, in order to satisfy the equations of motion, the flux combination
$G_3$ must be imaginary self-dual with respect to the Hodge operation
defined in terms of the Calabi-Yau metric in $X_6$
\beqa
*_6 \, G_3\, = \,i\, G_3
\label{isd}
\eeqa
These conditions guarantee the existence of a consistent supergravity 
solution for the different relevant fields in the configuration, metric, 
and 4-form, which have the form of a warped compactification (similar to 
a black 3-brane solution, since the same fields are sourced) 
\cite{dgs,gkp}. 

We remark that the condition (\ref{isd}) should not be regarded as an 
additional constraint on the fluxes. Rather, for a set of fluxes in a fixed 
topological sector (i.e. in a fixed cohomology class), it is a condition 
on the scalar moduli which determine the internal metric. The  scalar 
potential is minimized at points in moduli space where (\ref{isd}) is 
satisfied, while fluxes induce a positive scalar potential at other 
points. Hence introduction of fluxes leads to a natural mechanism to 
stabilize moduli. Explicit expressions will be discussed later on, for the 
moment let us state that generically the dilaton and all complex 
structure moduli are stabilized by this mechanism. On the other hand, 
Kahler moduli are not stabilized \cite{dgs,gkp,fp,kst}. 

\subsection{Supersymmetry}

The conditions for a configurations with 3-form fluxes to preserve some 
supersymmetry have been studied in \cite{gp}, and applied in explicit
constructions in \cite{fp,kst,kors}. 

The 10d $N =2$ type IIB real supersymmetry transformation parameters
$\epsilon_L$, $\epsilon_R$, can be gathered into a complex one
$\epsilon=\epsilon_L+i\epsilon_R$. It is chiral in 10d, satisfying
$\Gamma_{10d}\epsilon=-\epsilon$, with $\Gamma_{10d}=\gamma^0\ldots
\gamma^9$. Compactification on $X_6$ splits this spinor with respect 
to $SO(6)\times SO(4)$ e.g. as
\beqa
\epsilon_L=\xi\otimes u+\xi^*\otimes u^*
\eeqa
where $\xi$ is a 6d chiral spinor $\Gamma_{6d}\xi=-\xi$, and $u$ is a 
4d chiral spinor $\Gamma_{4d}u=u$. For $X_6$ of generic $SU(3)$
holonomy only one component of $\xi$ is covariantly constant and provides
susy transformations in 4d.

On the other hand, the presence of the O3-planes and D3-branes in the
background preserves only those $\epsilon$ satisfying
\beqa
\epsilon_R=-\gamma^4 \ldots \gamma^9 \epsilon_L
\eeqa
Such spinors are of the form $\epsilon=2\xi\otimes u$.

The conditions for a flux to preserve a supersymmetry associated to a 
particular spinor component of $\xi$ are \cite{gp}
\beqa
G\xi=0 \quad ; \quad G\xi^*=0 \quad ; \quad G\gamma^m \xi^*=0
\label{susyspinor}
\eeqa
where $G=\frac 16 G_{mnl} \gamma^{[m}\gamma^n\gamma^{l]}$.

Let us introduce complex coordinates $z^i,{\ov z}^i$ and define the 
highest weight state $\xi_0$ satisfying $\gamma^{\ov \i}\xi_0=0$. Then the 
O3-planes preserve $\xi_0$ and $\gamma^{ij}\xi_0$. Of these, a general 
Calabi-Yau (on which $z_i$ are complex coordinates) preserves only 
$\xi_0$, since it is $SU(3)$ invariant.

The conditions that a given flux preserves $\xi_0$, can be described 
geometrically \cite{gp,fp,kst} as

{\bf a)} $G_3$ is of type $(2,1)$ in the corresponding complex structure.

{\bf b)} $G_3$ is primitive, i.e. $G_3\wedge J=0$ where $J$ is the Kahler 
form.

For explicit discussion of these conditions see below. Notice that a $G_3$ 
flux which is not $(2,1)$ in a complex structure, may still be 
supersymmetric if it preserves other spinor $\xi_0'$ (although it does not 
preserve $\xi_0$). In such case, $G_3$ would be of type $(2,1)$ in a
different complex structure where $\xi_0'$ is the spinor annihilated by
the new $\gamma^{{\ov i}{'}}$. In a general Calabi-Yau, however, there is 
a prefered complex structure, so that a supersymmetric flux should obey 
the above conditions with respect to it.

Since the techniques to find consistent (possibly supersymmetric) fluxes
at particular values of the stabilized moduli (and vice versa) have been 
discussed in the literature, we will not dwelve into their discussion. 
 
\section{Chiral D-brane configurations}
\label{philosophy}

The models considered in \cite{kst,fp} succeed in leading to $N=1$ or 
non-supersymmetric low-energy theories, with stabilization of most 
moduli. However, they are unrealistic in that they are automatically 
non-chiral since the only gauge sectors live on parallel D3-branes whose 
low-energy spectra are non-chiral \footnote{The D3-brane world-volume
spectra are at best $N=1,0$ deformations of $N=4$ theories, by 
flux-induced operators breaking partially or totally the world-volume 
supersymmetry}.

We are interested in constructing models containing gauge sectors with 
charged chiral fermions, and a bulk with flux-induced moduli stabilization. 
There are several possibilities to do this, corresponding to the 
different ways to build configurations of D-branes containing chiral 
fermions. Equivalently, the different ways of enriching the simple 
configuration of parallel D3-branes, to lead to chiral open string 
spectra.

\medskip

{\bf D3-branes at singularities}

One possibility is to use compactification varieties containing singular 
points, e.g. orbifold singularities. Locating D3-branes at the singularity
leads to chiral gauge sectors, with low energy spectrum given by a quiver
diagram \cite{dm} \footnote{For phenomenological model building in this 
setup with no field strength fluxes, see \cite{singu}.}. A simple set of 
models can be constructed starting with type IIB theory on $\IT^6/\IZ_3$ 
modded out by the $\Omega R$ orientifold projection, where $\Omega$ is 
worldsheet parity and $R:z_i\to -z_i$. This is 
particularly promising, since it is the simplest orbifold which can lead 
to three families in the sector of D3-branes at singularities. However, 
it is not possible to obtain $N=1$ supersymmetric models in this setup, 
for the following reason. In the complex structure where the spinor 
invariant under $\IZ_3$ satisfies $\gamma^{\ov \i}\xi_0=0$, the $\IZ_3$ 
orbifold action reads
\beqa
\theta:(z_1,z_2,z_3)\to (e^{2\pi i/3} z_1, e^{2\pi i/3} z_2, e^{-4\pi 
i/3} z_3)
\eeqa
Fluxes preserving the same spinor $\xi_0$ should be of type $(2,1)$ in 
this complex structure, namely linear combinations of $d{\ov 
z}^1dz^2dz^3$, $dz^1d{\ov z}^2dz^3$ and $dz^1dz^2d{\ov z}^3$. 
Such fluxes are not invariant under the orbifold action, and cannot be
turned on. In other words, the only possible fluxes are not supersymmetric.
The same problem arises for other promising orientifolds, like
$\IZ_3\times \IZ_2\times \IZ_2$. 

So we will not pursue the construction of $N=1$ susy models in the setup 
of D3-branes at singularities, and center on non-supersymmetric models.
An amusing possibility is to build models where the D-brane sector 
preserves some supersymmetry, while the closed string sector is 
non-supersymmetric due to  the combination of fluxes and orbifold action. 
We provide an example of this kind in section \ref{zthree}.

\medskip

{\bf Magnetised D-branes}

Although it is not often explicitly stated, it is also possible to obtain 
chiral fermions from wrapped parallel D-branes, if the geometry of the 
wrapped manifold or the topology of the internal world-volume gauge 
bundle are non-trivial. The chiral fermions arise from the Kaluza-Klein 
reduction of the higher-dimensional worldvolume fermions if the index of 
the corresponding Dirac operator is non-zero.

The simplest such setup is type IIB compactified on $\IT^6$ (with an 
additional orientifold, and possibly orbifold projections), with 
configurations of D9-branes spanning all of spacetime (namely wrapped 
on $\IT^6$) with non-trivial gauge bundles on $\IT^6$. The simplest 
bundles are given by constant magnetic fields on each of the $\IT^2$,
so we call them magnetised D-brane configurations. They have been
considered in \cite{magnetised} in the absence of closed string field
strength fluxes. Magnetised D9-branes are related to D3-branes in that, 
due to world-volume Chern-Simons couplings, the non-trivial magnetic 
fields induce non-zero D3-brane charge on them (indeed, in the limit of 
very large magnetic fields, their charges reduce to those of a set of 
parallel D3-branes).

In the absence of fluxes, these configurations are related by T-duality to 
configurations of intersecting D6-branes, so that any model of the latter 
kind can be easily translated \cite{bgkl,rabadan} to a magnetised  
D9-brane setup. The advantage of using the magnetised D9-brane picture with 
O3-planes is that it is now straightforward to include NSNS and RR fluxes 
in the configuration, by applying the tools reviewed above (for the 
situation without D-branes). Notice that in the T-dual version of 
intersecting branes this corresponds to turning on a complicated set of
NSNS, RR and metric fluxes, see below. For the class of fluxes we 
consider, the picture of magnetised D9-branes is more useful.

In section \ref{magnetised} we review magnetised D9-brane configurations, 
first without NSNS and RR fluxes, and then discuss the introduction of the 
latter, and provide some explicit examples with interesting chiral gauge 
sectors.

{\bf Intersecting D6-branes}

Much progress has been made in D-brane model building using type IIA
D6-branes wrapped on intersecting 3-cycles in an internal space
(e.g. \cite{bdl,bgkl,afiru,bkl,imr,rest,susy,susy2}) \footnote{See 
\cite{orbif} 
for early work leading to non-chiral models, and \cite{reviews} for 
reviews}. 
However it is difficult to introduce NSNS and RR field strength 
fluxes in those setups; the difficulties can be seen from different 
perspectives. The models usually contain O6-planes, hence we need 
combinations of fluxes which source the RR 7-form. A possible combination 
is the RR 0-form field strength of type IIA (i.e. a cosmological constant 
of massive type IIA) and the NSNS 3-form field strength. This combination 
of fluxes has not been studied in the literature, so it is not a 
convenient starting point (see \cite{bc} for preliminary non-chiral 
results in this direction).

One may think that T-dualizing three times a model with O3-planes and RR
and NSNS fluxes would yield the desired configurations with O6-planes.
However, T-duality acts in a very non-trivial way on $H_3$, transforming 
some of its components into non-trivial components of the T-dual metric, 
which is no longer Calabi-Yau \cite{halflat,kstt}. The final 
configuration indeed would contain fluxes (RR and `metric fluxes') which 
source the RR 7-form, and would lead to Poincare invariant 4d models 
(consistently with T-duality). However, a full analysis of the T-dual 
geometry, and its properties (e.g. possible 3-cycles on which to wrap 
D6-branes) is lacking. 

Hence we will not attempt to discuss intersecting D-brane models with 
NSNS and RR fluxes.
 
\section{NSNS and RR fluxes in a type IIB orientifold with D3-branes at 
singularities.}
\label{zthree}

\subsection{D3-branes at singularities}

We would like to consider compactifications of type IIB theory, of the 
form $M_4\times X_6$, with  D3-branes located at singular points in 
the transverse space. Namely, their world-volume is $\IM_4\times P$, with 
$P$ a singular point in $\IX_6$. Since we are interested in the massless 
states in the open string spectrum, these are only sensitive to the local 
geometry around the singular point $P$, and are insensitive to the global 
structure of the compactification. Hence, it is enough to consider a 
local model for the geometry around $P$.

A simple class of singularities, on which string theory can be quantized 
exactly in $\alpha'$ (and hence, on which we know the open string spectrum 
on D3-branes at the singularity) are orbifold singularities. For instance, 
an orbifold singularity $\IC^3/\IZ_N$ is obtained from flat 6d space 
$\IR^6=\IC^3$ by identifying points related by the order $N$ action
\beqa
\theta: (z_1,z_2,z_3) \to (e^{2\pi i\,a_1/N} z_1, e^{2\pi i\,a_2/N} z_2, 
e^{2\pi i\,a_3/N} z_3)
\eeqa
which generates $\IZ_N$. Here $a_i\in\IZ$, and $\sum a_i=$ even, so that 
the space is spin. We will consider supersymmetry preserving geometries, 
which requries $\sum a_i\in 2N\IZ$. 

The origin in $\IC^3$ descends in the quotient to  a conical singular 
point, at which we locate the D3-branes.
The spectrum on a stack of $K$ D3-branes at the $\IC^3/\IZ_N$ singularity 
is obtained from the spectrum of $K$ D3-branes in flat space (which is 
4d $N=4$ supersymmetric $U(K)$ Yang-Mills), by keeping states invariant 
under the $\IZ_N$ action. In the latter, one should take into account the
(Chan-Paton) action of $\IZ_N$ on the gauge degrees of freedom, given by 
conjugation by a $K\times K$ matrix
\beqa
\gamma_{\theta} \, = \, \diag  \, (\, \id_{n_0}, e^{2\pi i/N} \id_{n_1},
\ldots, e^{2\pi i(N-1)/N} \id_{n_{N-1}} \, )
\eeqa
where $\sum_a n_a=K$. The $n_a$ are morevoer constrained by certain 
consistency conditions, the twisted RR tadpole cancellation conditions. A 
simple (but not always unique) choice satisfying them is $n_r=k$ and hence 
$K=Nk$. 

The computation of the spectrum has been discussed and gives the following 
$N=1$ supermultiplet content \cite{dm,singu}

\begin{center}
\begin{tabular}{cc}
$N=1$ Vector Multiplet & $\prod_a U(n_a)$ \\
$N=1$ Chiral Multiplet & $\sum_{a=1}^N \, [\, (\fund_a,\antifund_{a+a_1}) 
+ (\fund_a,\antifund_{a+a_2}) + (\fund_a,\antifund_{a+a_3})\, ]$ 
\end{tabular}
\end{center}
A concrete nice example is the $\IC^3/\IZ_3$ singularity, with 
$(a_1,a_2,a_3)=(1,1,-2)$, and which leads to a spectrum with a triplicated 
set of chiral multiplets (and hence of chiral fermions).
\begin{center}
\begin{tabular}{cc}
$N=1$ Vector Multiplet & $U(k)\times U(k)\times U(k)$ \\
$N=1$ Chiral Multiplet & $3 \, [\, (\fund_0,\antifund_1) \, + \,
(\fund_1,\antifund_2)\, + \, (\fund_2,\antifund_0)\, ]$ \\
\end{tabular}
\end{center}

For future convenience, we simply point out that for D3-branes at a 
singularity, mapped to itself under an orientifold action, an additional 
projection should be imposed. For such an orientifold of the $\IC^3/\IZ_3$ 
singularity, the spectrum is
\begin{center}
\begin{tabular}{cc}
$N=1$ Vector Multiplet & $SO(k-4)\times U(k)$ \\
$N=1$ Chiral Multiplet & $3 \, [\, (\fund,\antifund) \, + \,
(1,{\Yasymm}\, )\, ]$ \\
\label{ztres}
\end{tabular}
\end{center}

The analysis in this section applies both to models without and with 
fluxes. In the presence of the latter, the fluxes do not change the chiral 
structure of the theory, but may lead to additional operators in the 
D3-brane world-volume field theory (like e.g. gaugino masses, etc). These 
are computable in the regime of dilute fluxes, namely in the large volume 
limit \cite{soft}.

\subsection{Models with fluxes: A 3-family $SU(5)$ GUT example}

In this section we discuss the construction of a compactification with 
fluxes, and D3-branes at an orbifold singularity. Consider type IIB on 
$\IT^6$ modded by the orientifold action $\Omega R$, and quotient by the 
above $\IZ_3$ action 
\beqa
\theta: (z_1,z_2,z_3) \to (e^{2\pi i/3} z_1, e^{2\pi i/3} z_2, e^{-4\pi 
i/3} z_3)
\eeqa
This is a T-dual version of the model in \cite{abpss}. 

Now introduce NSNS and RR 3-form field strength fluxes. As 
discussed in previous section, the models turn out be non-supersymmetric in 
the closed string sector, since the orbifold and the $\IZ_3$ invariant 
fluxes necessarily preserve different supersymmetries.
Introduce a 3-form flux \footnote{In what follows we absorb the normalization factor $1/(4\pi^2 \alpha')^2$ appearing in \ref{Nflux} in the definition of the fluxes.} 
\beqa
G_3\, =\, 2\, d{\ov z}_1d{\ov z}_2d{\ov z}_3
\label{z3flux}
\eeqa
Such fluxes have been considered in ${\IT^6}/\Omega R$ in \cite{fp,kst}.
In this geometry all geometric moduli are Kahler, hence they are not 
stabilized. The above flux however, stabilizes the dilaton, which is fixed
at a value $\phi=e^{2\pi i/3}$.  The flux is imaginary self-dual with 
respect to the underlying metric. 
For the configuration to make sense it is crucial that the flux  
(\ref{z3flux}) is invariant under the action of $\theta$. It is also 
important to notice that the $\IZ_3$ quotient of $\IT^6/(\Omega R)$
does not contain closed 3-cycles which are not closed in $\IT^6/\Omega R$,
hence proper quantization is not spoilt. This is because the collapsed 
cycles at $\IZ_3$ singularities are 2- and 4-cycles, hence do not impose 
additional quantization constraints.

Notice that this flux preserves some of the supersymmetries of the
underlying ${\IT}^6/(\Omega R)$ geometry, namely the spinors 
$\gamma^i \xi_0$. These are however broken by the orbifold projection. 
It is possible that this kind of breaking of supersymmetry 
has some particularly nice features, since the interactions between 
untwisted modes is sensitive to supersymmetry breaking only via effects 
involving twisted modes. It would be interesting to analyze the impact of 
this property on the violations of the no-scale structure of the low 
energy supergravity effective theory for these models (i.e. the degree of 
protection against $\alpha'$ or $g_s$ corrections).

For the above flux we have $N_{\rm flux}=12$, hence cancellation of RR 
tadpoles requires the introduction of 20 D3-branes. To define the model 
completely, we need to specify the configuration of
these 20 D3-branes. In the $\Omega R$ orientifold of $\IT^6$ there is one 
point, the origin $(0,0,0)$, fixed under $\Omega R$ and $\theta$. At this 
point, cancellation of RR twisted tadpoles requires the presence of 
D3-branes, with a Chan-Paton matrix satisfying
\beqa
\Tr \, \gamma_{\theta}=-4
\eeqa
In addition, there are other $26$ points fixed under $\theta$ (and gathered
in 13 pairs under $\Omega R$), where there is no twisted RR tadpole. If
D3-branes are present, they should have traceless Chan-Paton matrix. Finally
there are 63 points fixed under $\Omega R$ (gathered in 21 trios under
$\IZ_3$) at which we may locate any number (even or odd) of D3-branes.

A simple solution would be to locate the 20 D3-branes at the origin, with
\beqa
\gamma_{\theta}=\diag (\id_4,e^{2\pi i/3} \id_8, e^{4\pi i/3} \id_8)
\eeqa
leading to an $N=1$ supersymmetric sector (to leading approximation,
since interactions with the closed sector would transmit supersymmetry
breaking), with spectrum (\ref{ztres})
\begin{center}
\begin{tabular}{cc}
$N=1$ Vector Multiplet & $SO(4)\times U(8)$ \\
$N=1$ Chiral Multiplet & $3 \, [\, (4,{\ov 8}) \, + \,(1,28)\, ]$ \\
\end{tabular}
\end{center}

A more interesting possibility, which we adapt from \cite{lpt}, is to
locate 11 D3-branes at the origin, with
\beqa
\gamma_{\theta}=\diag (\id_1,e^{2\pi i/3} \id_5, e^{4\pi i/3} \id_5)
\eeqa
This leads to a gauge sector with
\begin{center}
\begin{tabular}{cc}
$N=1$ Vector Multiplet & $U(5)$ \\
$N=1$ Chiral Multiplet & $3 \, [\, {\ov 5}+ 10 \, ]$ 
\end{tabular}
\end{center}

One should now be careful in locating the additional D3-branes. We have
introduced an odd number of D3-branes on top of the O3-plane at the
origin, and this implies that it is an $\widetilde{O3}^-$-plane in 
notation of \cite{witbar}, i.e. there exists a $\IZ_2$ $B_{RR}$ background 
on an ${\bf RP}_2$ around the O3-plane. In order to be consistent with the 
fact that our flux has even integral over the different 3-cycles implies 
that there should exist other $\widetilde{O3}^{-}$-planes in the 
configuration. The conditions in \cite{fp,kst} state that for any 3-plane 
over which the integrated flux of $H_3$ is even (any 3-plane in our case), 
the number of $\widetilde{O3}^{-}$-planes must be even. Here we consider a 
configuration, appeared in a model without fluxes \cite{lpt}, which 
satisfies this constraint.

Let us denote $A$, $B$ or $C$ the coordinate of an O3-plane in a complex 
plane, according to whether $z_i=1/2$, $z_i=e^{2\pi i/3}/2$ or 
$z_i=(1+e^{2\pi i/3})/2$. The remaining 9 D3-branes in the model are 
located on top of O3-planes at the points
\beqa
& (A,A,A)\quad (B,B,C) \quad (C,C,B) \nonumber \\
& (A,0,0) \quad (B,0,0)\quad (C,0,0) \nonumber \\
& (0,A,0) \quad (0,B,0) \quad (0,C,0)
\eeqa
This set is invariant under exchange of fixed points by $\IZ_3$, and
introduces the right number of $\widetilde{O3}^{-}$-planes at the right 
places. The additional D3-branes do not lead to additional gauge 
symmetries.

Thus the final model contains a 3-family $SU(5)$ GUT gauge sector (although
without adjoint chiral multiplets to break it down to the Standard Model),
as the only gauge sector of the theory. In addition, its closed string
sector is non-supersymmetric, but the cosmological constant vanishes at 
leading order. The impact of the supersymmetry breaking on the gauge 
sector is computable in the large volume limit \cite{soft}. 
It would also be interesting to construct other models based on the 
$\IZ_3$ orbifold, or other orbifold models. We leave these interesting 
question for future work.
 
\section{Magnetised D-branes}
\label{magnetised}

In this section we describe configurations of magnetised D9-branes, first 
before the introduction of fluxes, and then introducing them. We first 
consider the case of toroidal compactifications and orientifolds thereof. 
These models, in the absence of bulk fluxes, are T-dual to the models of 
intersecting D6-branes in toroidal compactifications \cite{afiru}, 
toroidal orientifolds \cite{bgkl} and orbifolds \cite{susy}. Subsequently 
we describe the introduction of fluxes and provide two explicit examples 
with the gauge symmetry and chiral fermion content of just the Standard 
Model.

\subsection{Magnetised D-branes in toroidal compactifications}
\label{toroidal}

We start with the simple case of toroidal compactification, with no 
orientifold projection. Consider the compactification of type IIB theory
on $\IT^6$, assumed factorizable.

We consider sets of $N_a$ D9-branes, labelled D9$_a$-branes, wrapped 
$m_a^i$ times on the $i^{th}$ 2-torus $({\IT}^2)_i$ in $\IT^6$, and with 
$n_a^i$ units of magnetic flux on $(\IT^2)_i$. Namely, we turn on a 
world-volume magnetic field $F_a$ for the $U(1)_a$ gauge factor in 
$U(N_a)$, such that
\beqa
m_a^i \, \frac 1{2\pi}\, \int_{\IT^2_{\,i}} F_a^i \, = \, n_a^i
\label{monopole}
\eeqa
Hence the topological information about the D-branes is encoded in the 
numbers $N_a$ and the pairs $(m_a^i,n_a^i)$ \footnote{Notice the change 
of roles of $n$ and $m$ as compared with other references. This however 
facilitates the translation of models in the literature to our language.} 

This description automatically includes other kinds of lower dimensional 
D-branes. For instance, a D7-brane (denoted D7$_{(i)}$) sitting at a 
point in $\IT^2_{\,i}$ and wrapped on the two remaining two-tori
(with generic wrapping and magnetic flux quanta) is described by 
$(m^i,n^i)=(0,1)$ (and arbritrary $(m^j,n^j)$ for $j\neq i$); 
similarly, a D5-brane (denoted D5$_{(i)}$) wrapped on $\IT^2_{\,i}$ (with 
generic wrapping and magnetic flux quanta) and at a point in the remaining 
two 2-tori is described by $(m^j,n^j)=(0,1)$ for $j\neq i$; finally, a 
D3-brane sitting at a point in $\IT^6$ is described by $(m^i,n^i)=(0,1)$ 
for $i=1,2,3$. This is easily derived by noticing that the boundary 
conditions for an open string ending on a D-brane wrapped on a two-torus 
with magnetic flux become Dirichlet for (formally) infinite magnetic 
field.

\medskip

D9-branes with world-volume magnetic fluxes are sources for 
the RR even-degree forms, due to their worldvolume couplings 
\beqa
\int_{D9_a} C_{10} \quad ;\quad
\int_{D9_a} C_{8}\wedge \tr F_a \quad ;\quad
\int_{D9_a} C_{6}\wedge \tr F_a^{\, 2} \quad ;\quad
\int_{D9_a} C_{4}\wedge \tr F_a^{\, 3} 
\eeqa
Consistency of the configuration requires RR tadpoles to cancel. Following the discussion in \cite{afiru}, leads to the 
conditions
\beqa
&\sum_a N_a m_a^1 m_a^2 m_a^3 = 0 \nonumber \\
&\sum_a N_a m_a^1 m_a^2 n_a^3 = 0 & {\rm and}\; {\rm 
permutations} \; {\rm of} \; 1,2,3\nonumber \\
&\sum_a N_a m_a^1 n_a^2 n_a^3 = 0 & {\rm and}\; {\rm 
permutations} \; {\rm of} \; 1,2,3\nonumber \\
&\sum_a N_a n_a^1 n_a^2 n_a^3 = 0 
\label{rrtadpoles}
\eeqa
Which amounts to cancelling the D9-brane charge as well as the induced 
D7-, D5- and D3-brane charges.

Introducing for the $i^{th}$ 2-torus the even homology classes 
$[{\bf 0}]_i$ and $[{\IT^2}]_i$ of the point and the two-torus, the vector 
of RR charges of one D9-brane in the $a^{th}$ stack is
\beqa
[{\bf Q}_a]\, =\, \prod_{i=1}^3\, (m_a^i [{\IT}^2]_i + n_a^i [{\bf 0}]_i) 
\label{charge}
\eeqa
The RR tadpole cancellation conditions (\ref{rrtadpoles}) read
\beqa
\sum_a \, N_a [{\bf Q}_a]\, = \, 0
\eeqa

The conditions that two sets of D9-branes with worldvolume magnetic 
fields
$F_a^i$, $F_b^i$ preserve some common supersymmetry can be derived from
\cite{bdl}. Indeed, it is possible to compute the spectrum of open strings
stretched between them and verify that it is supersymmetric if
\beqa
\Delta_{ab}^1 \pm \Delta_{ab}^2 \pm \Delta_{ab}^3 = 0
\eeqa
for some choice of signs (in order to preserve $\xi_0$, the choice should 
be all positive signs). Here
\beqa
\Delta_i=\arctan\, [(F_a^i)^{-1}]-\arctan \,[(F_b^i)^{-1}]
\eeqa
and
\beqa
F_a^i=\frac{n_a^i}{m_a^i R_{x_i} R_{y_i}}
\eeqa
which follows from (\ref{monopole}).

\medskip

The spectrum of massless states is easy to obtain. The sector of open 
strings in the $aa$ sector leads to $U(N_a)$ gauge bosons and 
superpartners with respect to the 16 supersymmetries unbroken by the 
D-branes. In the $ab+ba$ sector, the spectrum is given by $I_{ab}$ chiral 
fermions in the representation $(N_a,{\ov N}_b)$, where 
\beqa
I_{ab}=[{\bf Q}_a]\cdot [{\bf Q}_b]=\prod_{i=1}^3 (n_a^im_b^i-m_a^i n_b^i)
\label{zeromodes}
\eeqa
is the intersection product of the charge classes, which on the basic 
classes $[{\bf 0}]_i$ and $[{\IT}^2]_i$ is given by the bilinear 
antisymmetric form 
\beqa
\left ( \begin{array}{cc} 
0 & -1 \cr 1 & 0\end{array} \right )
\eeqa
The above multiplicity can be computed using the $\alpha'$-exact boundary
states for these D-branes \cite{bgkl}, or from T-duality with
configurations of intersecting D6-branes. We now provide an alternative
derivation which remains valid in more complicated situations where the
worldsheet theory is not exactly solvable. Consider for simplicity
a single two-torus. We consider two stacks of $N_a$ and $N_b$ branes
wrapped $m_a$ and $m_b$ times, and with $n_a$, $n_b$ monopole quanta.
Consider the regime where the two-torus is large, so that the magnetic 
fields are diluted and can be considered a small perturbation around the 
vacuum configuration. In the vacuum configuration, open strings within 
each stack lead to a gauge group $U(N_am_a)$ and $U(N_bm_b)$ respectively, 
which is subsequently broken down to $U(N_a)\times U(N_b)$ by the monopole 
background, via the branching
\beqa
U(N_a m_a)\times U(N_b m_b) \to U(N_a)^{m_a}\times U(N_b)^{m_b}\to 
U(N_a)\times U(N_b)
\label{branch1}
\eeqa
Open $ab$ strings lead to a chiral 10d fermion transforming in the 
bifundamental $(\fund_a,\antifund_b)$ of the original $U(N_a m_a)\times 
U(N_b m_b)$ group. Under the decomposition (\ref{branch1}) the 
representation splits as
\beqa
(\fund_a,\antifund_{\,b})\to ({\underline{\fund_a,\ldots}};{\underline{ 
\antifund_{\,b},\ldots}}) \to m_a m_b (\fund_a,\antifund_{\,b})
\label{branch2}
\eeqa
The 8d theory contains chiral fermions arising from these, because of the 
existence of a nonzero index for the internal Dirac operator (coupled to 
the magnetic field background). The index is given by the first Chern class
of the gauge bundle to which the corresponding fermions couples. Since it
has charges $(+1,-1)$ under the $a^{th}$ and $b^{th}$ $U(1)$'s, the
index is
\beqa
{\rm ind}\, \Dsl_{ab}\, = \, \int_{\IT^2}\, (F_a-F_b)= \frac{n_a}{m_a}
-\frac{n_b}{m_b}
\eeqa
Because of the branching (\ref{branch2}), a single zero mode of the Dirac
operator gives rise to $m_am_{\,b}$ 8d chiral fermions in the
$(\fund_a,\antifund_b)$ of $U(N_a)\times U(N_b)$. The number of chiral fermions
in the 8d theory in the representation $(\fund_a,\antifund_b)$ of the final
group is given by $m_a m_b$ times the index, namely
\beqa
I_{ab}= m_am_b\int_{\IT^2} (F_a -F_b) = n_a m_b-m_a n_b
\eeqa
The result (\ref{zeromodes}) is a simple generalization for the case of 
compactification on three two-tori.

Notice that the field theory argument to obtain the spectrum is valid only
in the large volume limit. However, the chirality of the resulting
multiplets protects the result, which can therefore be extended to
arbitrarily small volumes. This kind of argument will be quite useful in
the more involved situation with closed string field strength fluxes, where
we do not have a stringy derivation of the results.

\subsection{Magnetised D-branes in toroidal orientifolds}
\label{orienti}

We are interested in adding orientifold planes into this picture, since
they are required to obtain supersymmetric fluxes. Consider type IIB on
$\IT^6$ (with zero NSNS B-field) modded out by $\Omega R$, with $R:z_i\to
-z_i$. This introduces 64 O3-planes, which we take to be all $O3^-$. It 
also requires the D9-brane configuration to
be $\IZ_2$ invariant. Namely, for the $N_a$ D9$_a$-brane with topological
numbers $(m_a^i,n_a^i)$ we need to introduce their $N_a$ $\Omega R$ images
D9$_{a'}$ with numbers $(-m_a^i,n_a^i)$.

The RR tadpole cancellation conditions read
\beqa
\sum_a \, N_a [{\bf Q}_a]\, +\, \sum_a \, N_a [{\bf Q}_{a'}]\, - 32\,
[{\bf Q}_{O3}]\, =\, 0
\eeqa
with $[{\bf Q}_{O3}]= [{\bf 0}]_1\times [{\bf 0}]_2\times [{\bf 0}]_3$.
More explicitly
\beqa
&\sum_a N_a m_a^1 m_a^2 n_a^3 = 0 & {\rm and}\; {\rm 
permutations} \; {\rm of} \; 1,2,3\nonumber \\
&\sum_a N_a n_a^1 n_a^2 n_a^3 = 16 
\eeqa
Namely, cancellation of induced D7- and D3-brane charge. Notice that there 
is no net D9- or D5-brane charge, in agreement with the fact that the 
orientifold projection eliminates the corresponding RR fields 
\footnote{There is also an additional discrete constraint, which we skip 
for simplicity, see footnote 10 in \cite{cascur}.}

The rules to obtain the spectrum are similar to the above ones, with the 
additional requirement of imposing the $\Omega R$ projections. This 
requires a precise knowledge of the $\Omega R$ action of the different 
zero mode sectors. The analysis is simplest in terms of the
T-dual description, where it amounts to the geometric action of the
orientifold on the intersection points of the D-branes. The result, which
is in any case can be derived in the magnetised brane picture, is as 
follows \cite{bgkl}.

The $aa$ sector is mapped to the $a'a'$ sector, hence suffers no  
projection \footnote{We do not consider branes for which $a=a'$ here.}.  
We obtain a 4d $U(N_a)$ gauge group, and superpartners with respect to 
the $N=4$ supersymmetry unbroken by the brane.

The $ab+ba$ sector is mapped to the $b'a'+a'b'$ sector, hence does not 
suffer a projection. We obtain $I_{ab}$ 4d chiral fermions in the 
representation $(\fund_a,\antifund_b)$. Plus additional scalars which are 
massless in the susy case, and tachyonic or massive otherwise.

The $ab'+b'a$ sector is mapped to the $ba'+a'b$. It leads to $I_{ab'}$ 4d 
chiral fermions in the representation $(\fund_a,\fund_b)$ (plus additional 
scalars).

The $aa'+a'a$ sector is invariant under $\Omega R$, so suffers a 
projection. The result is $n_{\Yasymm}$ and $n_{\Ysymm}$ 4d chiral
fermions in the $\Yasymm_a$, $\Ysymm_a$ representations, resp, with
\beqa
n_{\Yasymm}= \frac 12(I_{aa'}+ 8I_{a,O3})=\,- 4 \,m_a^1 m_a^2 m_a^3 \,
(n_a^1n_a^2n_a^3+1) \nonumber \\
n_{\Ysymm}= \frac 12(I_{aa'}-8I_{a,O3})=\,- 4 \,m_a^1 m_a^2 m_a^3 \,
(n_a^1n_a^2n_a^3-1) 
\eeqa
where $I_{a,O3}=[{\bf Q}_a]\cdot [{\bf Q}_{O3}]$.

\subsection{A model with fluxes and D-branes, with just the Standard 
Model spectrum}

In this section we consider the introduction of fluxes, constructing two 
explicit models with the chiral content of the Standard Model.

\smallskip

Consider type IIB theory on $\IT^6$ and mod out by the orientifold action 
$\Omega R$. Let us introduce the flux.
\beqa
G_3\, =\, \frac{2}{\sqrt{3}}\, e^{-\pi i/6}\,
(\, d{\bar z}_1 dz_2 dz_3 \, + \,
dz_1 d{\bar z}_2 dz_3 \, + \, dz_1 dz_2 d{\bar z}_3 \,)
\eeqa
This flux \cite{kst} stabilizes the dilaton at $\phi=e^{2\pi i/3}$, and 
the $\IT^6$ geometry at a factorized product of three $\IT^2$, all with 
complex structure parameter $\tau=e^{2\pi i/3}$. The flux is manifestly 
$(2,1)$, and is supersymmetric for the subset of Kahler moduli satisfying 
the primitivity condition. The above choice of factorized geometry 
corresponds to $J=\sum A_i dz_i d{\ov z}_i$ and makes the flux 
supersymmetric.

The flux contributes to the 4-form tadpoles with $N_{\rm flux}=12$ units.
There is also the contribution of $-32$ arising from the 64 O3-planes. 

In order to cancel the remaining tadpole, we introduce a set of branes, 
which will contain the chiral gauge sector. We introduce $N_a$ 
D9$_a$-branes wrapping $m_a^i$ times on the $(\IT^2)_i$, and with $n_a^i$
units of world-volume magnetic flux on them.

In order to build interesting examples, we would like to choose a set 
of branes with the chiral spectrum of just the Standard Model
\footnote{There are two subtleties at this point. The first is that the 
set of consistent D-branes in a configuration with flux is classified by 
a modified K-theory group. The consistency of the configurations 
considered is discussed in appendix B in \cite{cascur}. The second is that 
our models contain D7-branes, whose moduli are also expected to be 
stabilized by the 3-form fluxes \cite{tt}. Hence, there may be 
additional restrictions, not taken into account, on the structure of the 
D7-brane stacks in our examples below.}. Sets of 
branes of this kind were described in \cite{imr}, mainly using the T-dual 
picture of D6-branes at angles. We may however use a suitable translation 
of their result. We take a configuration contributing 20 
units to the 4-form tadpole (and zero to others), for instance the choice
\beqa
\rho=1, \beta^1=\beta^2=1, \epsilon=1, n_a^2=n_d^2=2, n_b^1=1, n_c^1=4
\eeqa
in table 2 in \cite{imr}
\footnote{This requires introducing a non-zero $B_{NSNS}$ field in the 
third two-torus. This has been discussed in section 6 of \cite{cascur}.}. 
This leads to 

\begin{center}
\begin{tabular}{|c||c|c|c|}
\hline 
$N_a$ & $(m_a^1,n_a^1)$ & $(m_a^2,n_a^2)$ & $(m_a^3,n_a^3)$ \\
\hline\hline
$3$ & $(0,1)$ & $(1,2)$ & $(1/2,1)$ \\ 
\hline
$2$ & $(-1,1)$ & $(0,1)$ & $(3/2,1)$ \\ 
\hline
$1$ & $(3,4)$ & $(0,1)$ & $(1,0)$ \\ 
\hline
$1$ & $(0,1)$ & $(-1,2)$ & $(3/2,1)$ \\ 
\hline
\end{tabular}
\end{center}

This choice cancels all tadpoles. It leads to gauge group $U(3)\times 
U(2)\times U(1)^2$. It furthermore reduces to $SU(3)\times SU(2)\times 
U(1)$ after taking into account the $B\wedge F$ couplings; our choice of 
parameters is such that the $U(1)$ remaining massless precisely 
corresponds to hypercharge \footnote{Coupling of $B$ fields to closed 
string sector $U(1)$'s \cite{scp} could have modified the structure of the 
surviving $U(1)$, as compared with the situation in \cite{imr} (see also 
\cite{akr}). However, 
we have checked that the fields coupling to world-volume $U(1)$'s do not 
couple to closed string $U(1)$'s, so the condition for a massless 
hypercharge in \cite{imr} remains valid.}. Finally, it leads to chiral 
fermions multiplicites given by
\beqa
I_{ab}=1\, , \, I_{ab'}=2\, ,\, I_{ac}=-3\, ,\, I_{ac'}=-3 \, , \nonumber \\
I_{bd}=0\, ,\, I_{bd'}=-3\, ,\, I_{cd}=-3\, ,\, I_{cd'}=3
\label{smnumbers}
\eeqa
(others are zero), which leads to exactly the chiral spectrum of the 
Standard Model. It is important to point out that since $\prod_i m_a^i=0$, 
there are really no D9-branes, and the Wess-Zumino terms of 
\cite{urangaflux} are not present. The dynamical fermion content is 
non-anomalous.

The main features of this model are: The closed string sector is 
supersymmetric, while the open string sector is not. Open string tachyons 
are however avoided by choosing particular regions in the Kahler moduli 
space. Nevertherless, the model may lead to runaway potentials for Kahler 
moduli when the brane tension is taken into account, unless some 
Kahler moduli stabilization mechanism is included in the construction.

\medskip

There is another interesting kind of model that we would like to describe, 
and which is very similar in some respects to the models leading to 
deSitter vacua in string theory \cite{kklt}. The idea is to introduce too 
much flux, so that one overshoots the O3-plane RR tadpole, and to 
introduce a set of D-branes carrying net anti-D3-brane charge to cancel 
it. The extra contribution from the antibrane tension leads to a positive 
vacuum energy, which may turn the non-compact geometry into a deSitter 
space. This requires in addition a mechanism to stabilize the Kahler 
moduli, and to avoid a runaway behaviour instead of a deSitter vacuum, on 
which we will not enter. Hence our model is only reminiscent of 
\cite{kklt}, and not an explicit realization of their proposal. It is 
however interesting to describe configurations of this kind.

Let us consider a flux similar to the above, but with larger quanta
\beqa
G_3\, =\, 2\times \frac{2}{\sqrt{3}}\, e^{-\pi i/6}\,
(\, d{\bar z}_1 dz_2 dz_3 \, + \,
dz_1 d{\bar z}_2 dz_3 \, + \, dz_1 dz_2 d{\bar z}_3 \,)
\eeqa
This flux stabilizes moduli at the same values as the above one, but leads 
to a larger contribution to the RR 4-form tadpole, $N_{\rm flux}=48$ 
units.

In order to cancel the remaining tadpole, we introduce a set of branes, 
which will contain the chiral gauge sector, contributing with $-16$ units 
of RR 4-form tadpole. We choose another example from the general class in 
\cite{imr}, with
\beqa
\rho=1, \beta^1=-\beta^2=1, \epsilon=1, n_a^2=n_d^2=-1, n_b^1=n_c^1=2
\eeqa
in table 2 in \cite{imr}. This leads to 

\begin{center}
\begin{tabular}{|c||c|c|c|}
\hline 
$N_a$ & $(m_a^1,n_a^1)$ & $(m_a^2,n_a^2)$ & $(m_a^3,n_a^3)$ \\
\hline\hline
$3$ & $(0,1)$ & $(-1,-1)$ & $(1/2,1)$ \\ 
\hline
$2$ & $(-1,2)$ & $(0,-1)$ & $(3/2,1)$ \\ 
\hline
$1$ & $(3,2)$ & $(0,-1)$ & $(1,0)$ \\ 
\hline
$1$ & $(0,1)$ & $(1,-1)$ & $(3/2,1)$ \\ 
\hline
\end{tabular}
\end{center}

which cancels all RR tadpoles, and leads to a chiral sector with just the 
Standard Model spectrum, exactly as above.

\medskip

Very interestingly, our configuration corresponds to D7-branes with 
world-volume magnetic fields mimicking anti-D3-brane charge. This is a 
very explicit configuration realizing the proposal in \cite{quevedo} to 
replace antiD3-brane of \cite{kklt} by world-volume anti-instantons.
In particular, allows to address the absence of scalar vevs: there are 
regions in Kahler moduli space where no scalar tachyons are present, i.e. 
it is not possible to restabilize the supersymmetric vacuum since there 
are no scalars with the correct charges to cancel the FI term.

\medskip

Clearly, many other scenarios can be deviced. For instance one may break 
supersymmetry in the closed string sector, and consider a supersymmetric 
D-brane configuration, as in our example in section \ref{zthree}.
Hopefully our examples have provided a good illustration of the model 
building possibilities in this setup.

\subsection{Supersymmetric models?}

It is a natural question to wonder about the construction of 
supersymmetric chiral models with NSNS and RR fluxes. This has been 
attempted in \cite{blt,cascur} using orientifolds of 
$\IT^6/(\IZ_2\times \IZ_2)$ without success. The difficulty arises as 
follows: Chirality requires the introduction of non-trivial magnetic 
fields in the three two-tori, and this introduces several D-brane charges, 
which require several kinds of orientifold planes to cancel their charge 
in a supersymmetry-preserving way. In order to obtain several kinds of 
orientifold planes one needs orbifold quotients (since due to the group 
law the product of two orientifold actions is an orbifold action), 
with the above mentioned $\IZ_2\times \IZ_2$ being one of the simplest. 
The existence of orbifold projections usually modifies the flux 
quantization conditions (due to the requirement of proper quantization 
over cycles collapsed at the singularity, or similar subtleties), and 
requires flux quanta larger than in toroidal models. Such large fluxes 
generate a tadpole $N_{\rm flux}$ exceeding the negative value from the 
orientifold planes, so that tadpole cancellation requires the introduction 
of antibranes, which render the model non-supersymmetric. 

Although there is no general theorem in this direction, it seems difficult 
to construct supersymmetric chiral models with NSNS and RR fluxes using 
toroidal orbifolds. We expect that they however exist for orientifolds 
of more general Calabi-Yau manifolds, although such models would be more 
difficult to construct explicitly.
 
\section{Final comments}
\label{final}

Compactifications with field strength fluxes are a most promising avenue 
for model building, in that they provide a canonical mechanism to 
stabilize most moduli of the compactification. 

In this paper we have provided the basic model building rules for 
compactifications with D-branes, leading to chiral gauge sectors, and 
moduli stabilization by fluxes. We have described different 
approaches to achieve this aim, and provided explicit examples with 
D3-branes at singularities (with supersymmetry broken in the closed 
string sector) and D-branes with world-volume magnetic fluxes (with a 
supersymmetric closed string sector and a non-supersymmetric open string 
sector). 

Many further directions remain open. In the context of moduli stabilization, 
it would be interesting to explore the interplay between the flux induced 
scalar potential with other sources of potential for moduli, like 
non-supersymmetric sets of D-branes, or non-perturbative corrections. This 
step is crucial in order to understand the fate of the moduli which are 
not stabilized by the fluxes. It would also be interesting to determine 
the set of values at which moduli stabilize, in order to understand for 
instance what properties the underlying model must have in order to lead 
to e.g. small 4d gauge couplings, or large radii. 

Finally, our models contain several of the ingredients involved in the 
construction of deSitter vacua in string theory. It would be interesting 
to improve the kind of techniques discussed in the present paper, aiming 
towards building explicit models of this kind of constructions. This is 
essential in order to flesh out recent discussions on the discretuum of 
`realistic' vacua in string theory, and their implications for particle 
physics and cosmology.

\centerline{\bf Acknowledgements}

We thank G. Aldazabal, R. Blumenhagen, G. Honecker, L. Ib\'a\~nez, 
F. Marchesano, and F. Quevedo for fruitul conversations. A.M.U. thanks 
the organizers of the X Marcel Grossman meeting for an estimulating 
atmosphere at the meeting, and M.~Gonz\'alez for kind encouragement and 
support. J.G.C thanks M. P\'erez for her patience and affection. 

This work has been partially supported by CICYT (Spain). The research of J.G.C. is supported by the Ministerio de Educaci\'on, 
Cultura y Deporte through a FPU grant.

\newpage


\end{document}